\documentstyle[12pt]{article}
\begin{document}
\thispagestyle{empty}
\vspace*{-15mm}
\baselineskip 10pt
\begin{flushright}
\begin{tabular}{l}
{\bf OCHA-PP-199}\\
{\bf December 2002}\\
{\bf hep-th/0203223}
\end{tabular}
\end{flushright}
\baselineskip 24pt
\vglue 10mm

\begin{center}
{\LARGE\bf
Matrix Model on Z-Orbifold}
\vspace{7mm}

\baselineskip 18pt
{\bf
Akiko MIYAKE and Akio SUGAMOTO} 
\vspace{2mm} 

{\it
Department of Physics, Ochanomizu University}, \\
{\it 2-1-1, Otsuka, Bunkyo-ku, Tokyo 112-8610, Japan }\\

\vspace{10mm}
\end{center}
\begin{center}
{\bf Abstract}\\[7mm]
\begin{minipage}{14cm}
\baselineskip 16pt
\noindent
Six dimensional compactification of the type IIA matrix model on the ${\bf Z}$-orbifold is
studied. Introducing a ${\bf Z}_{3}$ symmetry properly on the three mirror images
of fields in the $N$-body system of the supersymmetric D0 particles,
the action of the Matrix model compactified on the ${\bf Z}$-orbifold is obtained.
In this matrix model ${\cal N}=1$ supersymmetry is explicitly
 demonstrated.

\end{minipage}
\end{center}

\newpage
\baselineskip 18pt
\def\thefootnote{\fnsymbol{footnote}}
\setcounter{footnote}{0}

\section*{Introduction}
Recent development of the string theory makes it possible to investigate
the strong coupling regime of the theory as well as the inter-relationship
between various string models (or phases) by the duality. To establish the
duality (exchange of ``electricity" and ``magnetism") it is necessary to
have a ``magnetic" object of string, in addition to the original "electric"
string. 

Once it is recognized by Witten~{\cite{Witten}} that the dynamics of the
extended objects,  D-branes~{\cite{Polchinski}}, playing an important role
in the duality, can be described in terms of  the dimensionally reduced
Supersymmetric Yang-Mills Theory (SYM), the non-perturbative study of string
becomes more familiar and
more realistic. In particular, the dynamics of the simplest point-like
D0 branes in 11 dimensions may play the fundamental role in such a
study~{\cite{BFSS}}. Since the coordinates of a gas, consisting of $N$
D0-branes, are represented by $N\times N$ hermitian matrices, its
dynamics is given in terms of the SYM theory dimensionally reduced to
one temporal dimension, that is the quantum mechanics of the SUSY Matrix
model. Hereafter, we refer this model simply as the Matrix model.

If the Matrix model is quantized in the light-cone gauge, the model 
reproduces the discrete version of the  super-membrane theory, since the 
discretized row and column indices $\sigma$ and $\rho$ of the Matrix,
$X^{\mu}(t)_{\sigma \rho}$, naturally become the coordinates of a membrane,
$X^{\mu}(t, \sigma, \rho)$ ~{\cite{Hoppe}}.   

Compactification of the Matrix model is an interesting topic, in particular
when the model is located on the orbifold space, {\it i.e.}, the space of
torus divided by its discrete symmetry.

First, since the 11d type IIA (yet-unknown) M-theory compactified on a
simplest orbifold space,  ${\bf S}^{1}/{\bf Z}_{2}$, is shown to be
connected with the 10d $E(8)
\times E(8)$ heterotic string theory~{\cite{Horava-Witten}}, from which
we obtain the well-known E(6) grand unified theory, after
compactification of 6d space by Calabi-Yau manifold
{\cite{Candelas}}. The Matrix model describing this (yet-unknown) M-theory
may be the type IIA Matrix model. 

Then, the compactification of this type IIA Matrix model on the orbifold
space ${\bf S}^{1}/{\bf Z}_{2}$ is demonstrated explicitly by
{\cite{Danielsson-Ferreti}}, {\cite{Kachra-Silverstein}}, {\cite{S.J.Rey}},
{\cite{Banks-Motl}}, and others. 

Second, when we intend to obtain ``realistic" 4d string models, by
compactifying the extra six dimensions in the 10d heterotic string models,
the orbifold compactification is a good method of breaking the
supersymmetries. Since the orbifold space does not differ so much from the
torus compactification, we can explicitly estimate various physical
quantities (amplitudes) in this orbifold compactification, without much
difficulty compared with the torus compactification. Among the orbifold
compactifications, we have found various realistic models with three
generations, having extra U(1) gauge groups and $\cal N$=1
supersymmetry~{\cite{Nilles}}, {\cite{Katsuki-Kobayashi}},
{\cite{Senda-Sugamoto}}. 

If the M-theory is compactified on a simple torus, ${\bf T}^{6}$, it is claimed
that the extra degrees of freedom, representing the wrapped D6 branes,
become massless and should be included, in order that the compactified
M-theory reproduces the strong coupling region of the corresponding type
IIA  superstring theory~\cite{T^6}.  Then, the theory becomes very
complicated, but the D6 brane may play the role of an ``electromagnetic
dual" of the D0 brane. These extra degrees of freedom coming from the
wrapped D6 branes may decouple in the other  6d compactification space,
such as the Calabi-Yau manifold~\cite{Kachra-Lawrence-Silverstein}.

In this paper, we study the second problem mentioned above, that is the six
dimensional compactification of the quantum mechanics of the type IIA
Matrix model by the {\bf Z}-orbifold~{\cite{Z-orbifold}}, the most
fundamental space among the various orbifold spaces.  Therefore, in our
study, we include only the D0-brane degrees of freedom (``magnetic" or
``electric" part), without incorporating the extra degrees of freedom
(``electric" or ``magnetic" part) existing in some cases of the M-theory.

If we consider our approach as the study of the type IIA Matrix model on
the non-compact orbifold, ${\bf C}^{3}/{\bf Z}^{3}$, then our obtained Matrix model may
become a candidate of M-theory on ${\bf C}^{3}/{\bf Z}^{3}$, since the wrapping of D6
branes is not possible in this case. 

The Matrix model on the ${\bf Z}$-orbifold is obtained from the $3N \times 3N$
Matrix model, by imposing the ${\bf Z}_3$ symmetry of the ${\bf
Z}$-orbifold, and ${\cal N}=1$ supersymmetry is explicitly checked in
the obtained model.

By generalizing our analysis to the more elaborate  orbifold spaces, we may
find the more realisitic analysis of string model on the basis of the
standard model with three generations of quarks and leptons. 

This work was presented by one of the authors (A. M.) at the JPS meetings
held at Niigata University in September 2000 and at Okinawa Kokusai
University in September 2001.

\section*{Matrix Model on Z-Orbifold}

We start with the type IIA Matrix model, {\it i.e.} the 10 dimensional Super Yang-Mills theory (SYM) dimensionally reduced to the single temporal dimension. The action of this model is written by 
\begin{eqnarray} 
S_{M}={\rm Tr}\int d\tau({\frac 1{2R}}(D_{\tau}X^{I})^{2} -{\frac R4}[X^{I}, X^{J}]^{2}
+\Theta^{T}D_{\tau}\Theta+iR\Theta^{T}\Gamma_{I}[X^{I}, \Theta]). 
\end{eqnarray}

Here, $I,J=0,\ldots ,9$ label the Minkowski space-time indices, $R$ is
																																																							    the radius of the compactified 11-th dimension, $\Gamma_{I}$ is the 10d gamma matrix, and $X^{I}$ and $\Theta$ are $3N\times 3N$ hermitian matrices, representing the gauge fields and the gaugino fields of the original SYM, respectively. The gaugino fields $\Theta$ are given by the Majorana-Weyl spinors, having eight degrees of freedom on the mass shell. In the terminology of the 3$N$-body D0 brane system, $X^{I}$ are the bosonic coordinates of the D0 branes. The $D_{\tau}$ is the covariant derivative defined by \begin{equation} D_{\tau} \equiv \partial_{\tau} - i[A_\tau, ~~]. \end{equation} 

Starting with this action, we study the action of the Matrix model
compactified on the {\bf Z}-orbifold of ${\bf T}^{6}/{\bf Z}_{3}$. For this purpose, it is
convenient to use the complex notations for the six spacial dimensions which we are going to compactify. By introducing complex coordinates $Z_{i},(i=1,2,3$),
\begin{eqnarray}
Z_{i}\equiv X^{2i+2}+iX^{2i+3} , \end{eqnarray} the torus lattice ${\bf
T}^{6}$ of the {\bf Z}-orbifold is defined by \begin{eqnarray}
					    Z_{i}\simeq Z_{i}+r \simeq Z_{i}+r e^{\pi i/3}, \label{torus lattice} \end{eqnarray} where $r$ is the compactification size of the six spacial dimensions.
This lattice has a ${\bf Z}_{3}$ symmetry, under \begin{eqnarray} Z_{i}\simeq e^{2\pi i/3}Z_{i}, \label{Z_{3} symmetry} \end{eqnarray}
for $i=1,\ldots ,3$, so that we can divide the torus lattice by this ${\bf Z}_{3}$ symmetry, and obtain the {\bf Z}-orbifold. 

Denoting uncompactified and compactified coordinates by $X_{_{//}}^{\mu}$ and $Z_{i}$, respectively ($\mu =0,\ldots ,3$ and $i=1,\ldots ,3$), where the complex coordinates $Z_{i}$ are assumed to satisfy the condition of the torus lattice (\ref{torus lattice}). 

We have to impose furthermore the ${\bf Z}_{3}$ symmetry (\ref{Z_{3} symmetry}) on the bosonic as well as the fermionic coordinates.
Under the ${\bf Z}_{3}$ symmetry, three points (or three mirror images) on the torus are identified. Corresponding to this identification we have to prepare three copies (or three mirror images) of the fields, and identify them by the ${\bf Z}_{3}$ symmetry up to the complex phases. Therefore, we start from the $3N\times 3N$ matrices for the $N$-body system of the D0 branes. 

Now the ${\bf Z}_{3}$ invariance we impose on the bosonic and fermionic fields are given as follows: \begin{eqnarray}
X_{_{//}}^{\mu}&=&MX_{_{//}}^{\mu}M^{\dag},\\ Z_{i}&=&\alpha_{i}MZ_{i}M^{\dag}, \label{Field Transformation 1} \\ \Theta&=&{\hat \alpha}M\Theta M^{\dag}, \label{Field Transformation 2} \end{eqnarray}
where the matrix $M$ is the generator of the ${\bf Z}_{3}$ transformation on the
fields, and it satisfys $M^{3}={\bf 1}_{3N\times 3N}$. The complex phases $\alpha_{j}$ and ${\hat \alpha}$ for the bosonic and fermionic fields respectively appearing under the ${\bf Z}_{3}$ transformation also satisfy $M^{3}=1$ and $(\alpha_{j})^{3}=({\hat \alpha})^{3}=1$. Then, $M$, $\alpha_{j}$ and ${\hat \alpha}$ can be written as follows: \begin{eqnarray} M&=&
\left(
\begin{array}{ccc}
0&e^{i\phi_{3}}&0\\
0&0&e^{i\phi_{1}}\\
e^{i\phi_{2}}&0&0
\end{array}
\right),\\
\alpha_{j}&=&{\rm exp}(2\pi i \frac{n_{j}}3),\\ {\hat \alpha}&=&{\rm exp}(2\pi i \sum_{j=2}^{4}\frac{n_{j}}3 b_{j}^{\dag} b_{j}), \label{spinor rotation}
\end{eqnarray}
where $n_{j}$ is an integer and a relation, $\phi_{1}+\phi_{2}+\phi_{3}=2\pi n$ ($n$= integer), is assumed to hold.
In Eq. (\ref{spinor rotation}), there appear the raising and lowering operators of the five ``spins" constituting the 10d spinor. These are defined, as usual, by
\begin{eqnarray}
\left\{
\begin{array}{l}
b_{0}=\frac 12 (\Gamma^{1}-\Gamma^{0})\\ b_{j}=\frac 12 (\Gamma^{2j}-i\Gamma^{2j+1})~,\quad j=1,\ldots ,4 \end{array}\right. \end{eqnarray}
and $\Gamma^{\mu}$ satisfy \begin{eqnarray} \{ \Gamma^{\mu},\Gamma^{\nu} \}=2\eta^{\mu\nu}\nonumber\\ ( \Gamma^{0})^{2}=-1~,\quad (\Gamma^{i})^{2}=1. \end{eqnarray}
Then, we can understand in Eqs.(\ref{Field Transformation 1}) and (\ref{Field Transformation 2}) that the ${\bf Z}_{3}$ rotation of the complex coordinates induces exactly that of the spinors by ${\hat \alpha}$, and the matrix $M$ interchanges cyclically the 3 ``mirror" images of the bosonic and fermionic fields.

The 10d spinor on the mass shell, can be expressed as a set of \{$\Theta_{(-, \pm, \pm, \pm, \pm)}$ \}, where + means the ``spin" at the position $j=0,\ldots,4$ is occupied, while - means the ``spin" is unoccupied. Due to the Majorana condition $B\Theta=\Theta^{*}$, with $B=\Gamma^{3}\Gamma^{5}\Gamma^{7}\Gamma^{9}$, we have the following relation:
\begin{equation}
\Theta_{(-,s_{2},s_{3},s_{4},s_{5})}^{*} = (s_{2}s_{3}s_{4}s_{5}) \Theta_{(-,-s_{2},-s_{3},-s_{4}, -s_{5})}, \end{equation} where $s_{j}=\pm$.
Then, we can decompose the Majorana fermion $\Theta$ into 3 components,
$\Theta_{1}$, $\Theta_{\omega}$, and $\Theta_{\omega^{2}}$, depending on
the eigen-value $\alpha=1, \omega, \omega^{2}$ for the operation ${\hat
\alpha}$, respectively, where $\omega =\exp (2\pi i/3)$.

We can recognize that there are three cases for ${\hat \alpha}$, namely, \begin{eqnarray}
(n_{2},n_{3},n_{4})=(1,1,1),~(1,1,2),~(1,2,2). \end{eqnarray} In the
case of $(n_{2},n_{3},n_{4})=(1,1,1)$, $\Theta_{1}$ is one (Majorana) 4d
spinor whose charge conjugated field is included within itself,
$\Theta_{\omega}$ are the 3 component
 4d spinors, and $\Theta_{\omega^{2}}$ are the charge conjugation of $\Theta_{\omega}$. These $\Theta_{1}$, $\Theta_{\omega}$, and $\Theta_{\omega^{2}}$ transform \{$1$\}, \{$3$\}, and \{$3^{*}$\} under the flavor SU(3) symmetry which corresponds to the rotational symmetry in the 3d complex spaces with axes $Z_{i} (i=1, \ldots, 3).$
In the $(1,1,2)$ case $\Theta_{\omega}$ are the charge conjugation of
$\Theta_{1}$, and $\Theta_{\omega^{2}}$ are invariant under the charge
conjugation, while in the $(1,2,2)$ case, $\Theta_{1}$ are the charge
conjugation of $\Theta_{\omega^{2}}$, and $\Theta_{\omega}$ are invariant
under the charge conjugation.

In the following, we adopt the case of $(1,1,1)$, for simplicity. 
Then, one finds that $3\times 3$ block matrix structure of the 
uncompactifited transverse coordiantes, the compactified coordiantes, 
and  the fermionic variables reads
\begin{eqnarray}
X_{_{//}}^{\mu}&=&
\left(
\begin{array}{ccc}
H&Ae^{i\phi_3}&A^{\dag}e^{-i\phi_2}\\
A^{\dag}e^{-i\phi_3}&H&Ae^{i\phi_1}\\
Ae^{i\phi_2}&A^{\dag}e^{-i\phi_1}&H
\end{array}
\right)_{3N\times 3N}^{\mu}\\
Z_{i}&=&
\left(
\begin{array}{ccc}
B_{1i}&\alpha_i^{-1}e^{i\phi_3}B_{2i}&\alpha_i^{-2}e^{-i\phi_2}B_{3i}\\
e^{-i\phi_3}B_{3i}&\alpha_i^{-1}B_{1i}&\alpha_i^{-2}e^{i\phi_1}B_{2i}\\
e^{i\phi_2}B_{2i}&\alpha_i^{-1}e^{-i\phi_1}B_{3i}&\alpha_i^{-2}B_{1i}
\end{array}
\right)_{3N\times 3N}\\ \Theta&\equiv&\sum_{a}\Theta^{a}T^{a}\\
T^{a}&=&
\left\{
\begin{array}{rll}
&\left(
\begin{array}{ccc}
{\hat H}&{\hat A}e^{i\phi_3}&{\hat A}^{\dag}e^{-i\phi_2}\\
{\hat A}^{\dag}e^{-i\phi_3}&{\hat H}&{\hat A}e^{i\phi_1}\\
{\hat A}e^{i\phi_2}&{\hat A}^{\dag}e^{-i\phi_1}&{\hat H}
\end{array}
\right)_{3N\times 3N}^a \quad &\mbox{for ~$\Theta$~with~$\alpha=1$},\\ 
&\left( \begin{array}{ccc}
{\hat B}_{1}&\alpha^{-1}e^{i\phi_3}{\hat B}_2&\alpha^{-2}e^{-i\phi_2}{\hat B}_3\\
e^{-i\phi_3}{\hat B}_3&\alpha^{-1}{\hat B}_1&\alpha^{-2}e^{i\phi_1}{\hat B}_2\\
e^{i\phi_2}{\hat B}_2&\alpha^{-1}e^{-i\phi_1}{\hat B}_3&\alpha^{-2}{\hat B}_1
\end{array}
\right)_{3N\times 3N}^{a} \quad &\mbox{for~$\Theta$~with $\alpha=\omega$, $\omega^{2}$}.
\end{array}
\right.
\end{eqnarray}
where
\begin{eqnarray}
\Omega_j\equiv
\left(
\begin{array}{ccc}
1& & \\
&\alpha_j^{-1}& \\
& &\alpha_j^{-2}
\end{array}
\right).
\end{eqnarray}
In the above expressions, $H$ and ${\hat H}$ stand for arbitrary
hermitian $N\times N$ matrices, and $A,~B_{1i},~B_{2i},~B_{3i},~{\hat
A},~{\hat B_1},~{\hat B_2}$ and ${\hat B_3}$ stand for arbitrary 
complex $N\times N$ matrices. 

The original action of the type IIA Matrix model compactified on the {\bf Z}-orbiforld is now deformed to
\begin{eqnarray}
S&=&{\rm Tr}\int d\tau \left\{ \frac 1{2R}(D_{\tau} X_{_{//}}^{\mu})^{2} +\frac 1{2R} (D_{\tau} Z_{i})(D_{\tau} Z_{i}^{\dag}) -\frac R4 \left([X_{_{//}}^{\mu}, X_{_{//}}^{\nu}]^{2} \right.\right.\nonumber\\ &&\left.\left.+ 2[X_{_{//}}^{\mu},Z_{i}][X_{_{//}}^{\mu}, Z_{i}^{\dag}] +\frac 12 [Z_{i}, Z_{j}][Z_{i}^{\dag}, Z_{j}^{\dag}]+ \frac 12 [Z_{i}, Z_{j}^{\dag}][Z_{i}^{\dag}, Z_{j}]\right)\right.\nonumber\\ &&\left.+\bar{\Theta}iD_{\tau}\Theta+R(\bar{\Theta} \Gamma_{i}[X_{_{//}}^{i}, \Theta]+\bar{\Theta}b_{i+1}[Z_{i}, \Theta]+ \bar{\Theta}b_{i+1}^{\dag}[Z_{i}^{\dag}, \Theta])\right\} . \end{eqnarray}

Here we impose the gauge fixing condition under the background field of $X^{0}_{cl}$, that is 

\begin{eqnarray}
\partial_{\tau}X^{0}- i[X^{i}_{ cl},X^{i}]=0\quad,(i=1,\ldots ,9). \end{eqnarray}

By decomposing the $3N \times 3N$ matrices into the sum of the tensor products ($N\times N$ matrices) $\otimes$ ($3 \times 3$ matrices), we have \begin{eqnarray}
X_{_{//}}^{\mu}&=&
H^{\mu}\otimes {\bf 1}_{3}+A^{\mu}\otimes M+A^{\mu\dag}\otimes M^{\dag},\\ 
Z_{i} &=&(B_{1i}\otimes {\bf 1}_3+B_{2i}\otimes M +B_{3i} \otimes M^{\dag})
\Omega_i ,\\
\Theta
&=&\Theta_{1}^{a}({\hat H}_{1}^{a}\otimes {\bf 1}_{3}+{\hat A}_{1}^{a} \otimes M+{\hat A}_{1}^{a\dag}\otimes M^{\dag}) \nonumber\\
\quad&\quad&+\Theta_{\omega}^{a}({\hat B}_{1\omega}^a\otimes {\bf 1}_3
+{\hat B}_{2\omega}^a\otimes M+{\hat B}_{3\omega}^a\otimes M^{\dag})
\Omega\nonumber\\ 
\quad&\quad&+\Theta_{\omega^{2}}^{a}({\hat B}_{1\omega^2}^a\otimes {\bf 1}_3
+{\hat B}_{2\omega^2}^a\otimes M+{\hat B}_{3\omega^2}^a\otimes M^{\dag})
\Omega^{\dag},\\
C&=&C_{gh}\otimes {\bf 1}_3 +A_{gh}\otimes M+A_{gh}^{\dag}\otimes M^{\dag},\cr \bar{C}&=&C_{gh}^{\dag}\otimes {\bf 1}_3 +A_{gh}^{\dag}\otimes M^{\dag} +A_{gh}\otimes M,
\end{eqnarray}
where $C$ and $\bar{C}$ are $3N\times 3N$ ghost and anti-ghost fields, $C_{gh}$ is the anti-commuting $N\times N$ hermitian matrix and $A_{gh}$ is the anti-commuting $N\times N$ arbitrary complex matrix. 

Performing the trace with respect to $3\times 3$ matrices, the bosonic
action, the fermionic action and the ghost action of our model are
obtained. The resulting actions are given in Eqs. 
(\ref{bosonic action}),~(\ref{fermionic action}) and (\ref{ghost
action}) in the appendix, where $\Omega_i =\Omega$.

This is the action of the type IIA Matrix model in which the six spacial dimensions are compactified on the {\bf Z}-orbifold. The effective action of this model in the presence of the classical configuration of $N$ D0-branes, and  the analysis of the Matrix model being compactified on the spaces ${\bf S}^{1}/{\bf Z}_{2} \times {\bf T}^{6}/{\bf Z}_{3}$, will be given in our forthcoming paper~{\cite{Miyake-Sugamoto}}.

Finally, we study the structure of supersymmetries. 
More precisely, for our choice
of $(n_{2},n_{3},n_{4})=(1,1,1)$, the isotopic structure of the bosonic
variable, $X^{\mu}_{_{//}}$, and the fermionic variable, $\Theta_{1}$,
is the same, both expressed in terms of $H$ and $A$, so that a naive
counting gives $3N^{2}$ degrees of freedom to both of them. After a
usual gauge fixing, two degrees of freedom among the vector
fields,  $X^{\mu}_{_{//}}$, remain to be physical, and the fermionic
field, $\Theta_{1}$, on the mass shell, contains two \{1\}'s under the
flavor SU(3) symmetry. Hence, the bosonic and fermionic degrees of
freedom for the uncompactified dimensions are the same number,
$6N^{2}$. As for the pair of bosonic fields, $Z^{i=1,\ldots ,3}$, and
the fermionic fields, $\Theta_{\omega}$, the isotopic structure of both
of them are expressed by $B_{1i},~B_{2i},~B_{3i},~{\hat B_1},~{\hat
B_{2}}$ and ${\hat B_3}$, giving $2N^{2}$ degrees of freedom. The degrees of freedom
coming from the Lorents structure gives the factor three to
$Z^{i=1,\ldots ,3}$, while the same factor three exists in
$\Theta_{\omega}$ on the mass shell, being attributable to $\{3\}$ under the flavor SU(3) symmetry. The same counting holds for the other pair of $Z^{i=1,\ldots ,3~\dagger}$ and $\Theta_{\omega^{2}}$. Hence, the bosonic and fermionic degrees of freedom in the compactified dimensions are both $18N^{2}$. 

Therefore, the uncompactified dimensions give $6N^{2}$ vector multiplets of (gauge bosons, gauginos), while the compactified dimensions give $18N^{2}$ chiral multiples of (higgs, higgsinos). 

Our starting action is that of SYM, which is invariant under the following supersymmetry, generated by the generator $Q$ and its spinorial transformation parameter $\epsilon$:
\begin{equation}
{\bar \epsilon}Q =-{\rm Tr}\left\{\left({\bar \epsilon}\Gamma^I \Theta\right) \left(D_\tau X_I\right)+\left(2\Sigma^{0I}D_\tau X_I -R\Sigma^{IJ} [X_I,X_J]\right)\epsilon_{\alpha}\Theta_\alpha^{\dag}\right\} ,
\end{equation}
where $\Sigma^{IJ}=1/4[ \Gamma^{I}, \Gamma^{J} ]$.

After decomposing the bosonic and fermionic fields as well as the spinorial parameter $\epsilon$ into the various components, following the analysis of the ${\bf Z}$-orbifold compactification given above, we find the following expression for the terms of ${\bar \epsilon}Q$, having time-derivative of the bosonic fields,  or the terms with the conjugate momenta of the bosonic fields:
\begin{eqnarray}
{\bar \epsilon}_{\alpha}{\rm Tr}\left(\Gamma^I_{\alpha\beta}{\dot X}_I
\Theta_\beta\right)
=3\epsilon_s^{*[234]} {\rm Tr}\left({\dot H}^{2+i3}{\hat H}+{\dot A}^{2+i3}{\hat A}^{\dag}+({\dot A}^{2-i3})^{\dag}{\hat A}\right)
\psi_s^{[1234]}\cr
+3\epsilon_s^{*[1]} {\rm Tr}\left({\dot H}^{2-i3}{\hat H}+{\dot A}^{2-i3}{\hat A}^{\dag}+({\dot A}^{2+i3})^{\dag}{\hat A}\right)\psi_s \cr
-3\epsilon_s^{*[i]} {\rm Tr}\left({\dot B}_{1j}{\hat B}_{1\omega^2} 
+\alpha^{-1}{\dot B}_{2j}{\hat B}_{3\omega^2}+\alpha {\dot B}_{3j}
{\hat B}_{2\omega^2}\right)\psi_s^{[ij]}\cr
-3\epsilon_s^{*[1jk]} {\rm Tr}\left({\dot B}_{1j}^{\dag}{\hat B}_{1\omega}
+{\dot B}_{2j}^{\dag}{\hat B}_{2\omega}+{\dot B}_{3j}^{\dag}
{\hat B}_{3\omega}\right)\psi_s^{[1k]},
\end{eqnarray}
where ${\dot H}^{2+i3}={\dot H}^2 +i{\dot H}^3$, and so on.

Here, $\epsilon$ and $\Theta$ are 10d Majorana-Weyl fermions, so that their components are defined as follows:
\begin{eqnarray}
{\bar \epsilon}=<0|b_1\epsilon_s^{*[1]}+<0|b_4 b_3 b_2\epsilon_s^{*[234]}
+\sum_{i=2}^{4}<0|b_i\epsilon_s^{*[i]} +\frac{1}{2}
\sum_{i,j=2}^{4}<0|b_j b_i b_1\epsilon_s^{*[1ij]}\\ 
\Theta =\psi_s |0>+\psi_s^{[1234]}b_1^{\dag}b_2^{\dag}b_3^{\dag}b_4^{\dag}|0>
+\sum_{i=2}^{4}\psi_s^{[1i]}b_1^{\dag}b_i^{\dag}|0>+\frac{1}{2}
\sum_{i,j=2}^{4}\psi_s^{[ij]}b_i^{\dag}b_{j}^{\dag}|0>,
\end{eqnarray}
with the Majorana conditions of $\epsilon_s^{*[1]} =\epsilon_s^{[234]}$,
$\epsilon_s^{*[i]}=-\frac{1}{2}\epsilon_{1ijk}\epsilon_s^{[1jk]}$, and
$\psi_s^* =\psi_s^{[1234]}$,
$\psi_s^{*[ij]}=-\frac{1}{2}\epsilon_{ijkl}\psi_s^{[kl]}$.

Now, we find that the bosonic field $Z^{1}_{//}$ is transformed to the
fermionic field $(\Theta_{1})_{s}$, by the SUSY transformation with the
parameter $\epsilon_s^{[234]}$, while the bosonic field $B_{i}$ is
transformed to the fermionic field  $(\Theta_{\omega^2})^{[ij]}_{s}$ by the
SUSY transformation with the parameter $\epsilon_{s}^{[j]}$. 
When $\Omega_i =\Omega^{-1}$, $B_i$ is transformed to
$(\Theta_{\omega})_s^{[1i]}$. 
In other words, the SUSY transformation obtained here for the whole
fields in our model, is defined by restricting
the transformation parameter $\epsilon$ as 
\begin{eqnarray}
\epsilon _s^{[234]}=\epsilon_s^{[i]}=\epsilon_s^{*[1]}=-\frac{1}{2}
\epsilon_{1ijk}\epsilon_s^{*[1jk]}\equiv\eta,
\end{eqnarray}
where $\eta$ is a common SUSY transformation parameter, expressed in
terms of a 4d chiral fermion on the mass shell. Hence, the $\cal N$=1
SUSY is explicitly demonstrated.  

In this model, $(X_{//}^{\mu}, \Theta_1)$ can be considered as a vector
multiplet, while $(Z^i, \Theta_{\omega^2})$ can be three chiral
multiplets. If these three chiral multiplets correspond to three
generations of quarks and leptons, this  $Z_{3}$ orbifold model
becames more interesting. In order to clarify the existing gauge group
of our model, we have to follow the steps given in
{\cite{Kachra-Silverstein}} by demonstrating
furthermore $Z_2$ orbifoldong and S-duality transformation.

The number of supersymmetry depends on the compactified
spaces. Different choices of the 6d orbifold spaces may lead to a
different analysis of the 4d Matrix models.

\section*{Acknowledgment}
The authors are grateful to Madoka Nishimura, Tadahito Nakajima and Kenji Suzuki for the useful discussions and the initial collaboration in this work. 
One of the authors (A.M.) gives her thanks to S. Iso for the valuable
discussion at the KEK workshop this March, where S. Iso gave a talk on
the ${\bf Z}_3$ Orbifolding of the IIB Matrix model carried out by his group~\cite{Iso}.

\section*{Appendix}
{\footnotesize 
\begin{eqnarray}
&S&(\mbox{bosonic action})\nonumber\\
&=&3{\rm Tr}\int d\tau\left[\frac 1{2R}\left\{\left({\dot H}^i
-i([H_{0},H^{i}]+ [A_{0},A^{i\dag}]+[A_{0}^{\dag},A^{i}])\right)^{2}
\right.\right.\nonumber\\ 
&&\left.\left.+2\left({\dot A}^{i}-i([H_{0},A^{i}]+[A_{0},H^{i}] 
+[A_{0}^{\dag},A^{i\dag}]) \right)\right.\right.\nonumber\\
&&\quad\left.\left.\left({\dot A}^{i\dag}-i([H_0,A^{i\dag}]
+[A_0,A^i]+[A_{0}^\dag ,H^i])\right)\right.\right.\nonumber\\ 
&&\left.\left.\left({\dot B}_{1i}-i([H_0,B_{1i}]+(A_0B_{3i}
-\alpha B_{3i}A_0)+(A_0^\dag B_{2i}-\alpha^{-1}B_{2i}A_{0}^\dag ))
\right)\right.\right.\nonumber\\
&&\left.\left.\left({\dot B}_{1i}^{\dag}-i([H_0,B_{1i}^{\dag}]
+(\alpha A_0 B_{2i}^{\dag}-B_{2i}^{\dag}A_0)+(\alpha^{-1}A_0^{\dag} 
B_{3i}^{\dag}-B_{3i}^{\dag}A_0^{\dag}))\right)\right.\right.\nonumber\\
&&+\left.\left.\left({\dot B}_{2i}-i([H_0,B_{2i}]+(A_0 B_{1i}
-\alpha B_{1i}A_0)+(A_0^\dag B_{3i}-\alpha^{-1}B_{3i}A_0^{\dag}))
\right)\right.\right.\nonumber\\
&&\left.\left.\left({\dot B}_{2i}^{\dag}-i([H_0,B_{2i}^{\dag}]
+(\alpha A_0 B_{3i}^{\dag}-B_{3i}^{\dag}A_0)+(\alpha^{-1}
A_0^{\dag}B_{1i}^{\dag}-B_{1i}^{\dag}A_0^{\dag}))\right)
\right.\right.\nonumber\\
&&+\left.\left.\left({\dot B}_{3i}-i([H_0,B_{3i}]+(A_0 B_{2i}
-\alpha B_{2i}A_0)+(A_0^{\dag}B_{1i}-\alpha^{-1} B_{1i}A_0^{\dag}))
\right)\right.\right.\nonumber\\
&&\left.\left.\left({\dot B}_{3i}^{\dag}-i([H_0,B_{3i}^{\dag}]
+(\alpha A_0 B_{1i}^{\dag}-B_{1i}^{\dag}A_0)+(\alpha^{-1}
A_0^{\dag}B_{2i}^{\dag}-B_{2i}^{\dag}A_0^{\dag}))\right)\right\}
\right.\nonumber\\
&&\left.-\frac R4\left\{\left([H^{i},H^{j}]+[A^{i},A^{j\dag}]
+[A^{i\dag},A^{j}]\right)^{2} \right.\right.\nonumber\\
&&\left.\left.+2\left([H^{i},A^{j}]+[A^{i},H^{j}]+[A^{i\dag},A^{j\dag}]\right) 
\left([H^{i},A^{j\dag}]+[A^{i\dag},H^{j}]+[A^{i},A^{j}]\right)
\right.\right.\nonumber\\
&&\left.\left.+2\left([H^i,B_{1j}]+(A^i B_{3j}-\alpha B_{3j}A^i)
+(A^{i\dag} B_{2j}-\alpha^{-1}B_{2j}A^{i\dag})\right)\right.\right.\nonumber\\
&&\left.\left.\left([H^i,B_{1j}^{\dag}]+(\alpha A^i B_{2j}^{\dag}
- B_{2j}^{\dag}A^i)+(\alpha^{-1}A^iB_{3j}^{\dag}-B_{3j}^{\dag}A^{i\dag})\right)\right.\right.\nonumber\\
&&\left.\left.+2\left([H^i,B_{2j}]+(A^i B_{1j}-\alpha B_{1j}A^i)
+(A^{i\dag} B_{3j}-\alpha^{-1}B_{3j}A^{i\dag})\right)\right.\right.\nonumber\\
&&\left.\left.\left([H^i,B_{2j}^{\dag}]+(\alpha A^i B_{3j}^{\dag}
-B_{3j}^{\dag}A^i)+(\alpha^{-1}A^{i\dag}B_{1j}^{\dag}-B_{1j}^{\dag}
A^{i\dag})\right)\right.\right.\nonumber\\
&&\left.\left.+2\left([H^i,B_{3j}]+(A^iB_{2j}-\alpha B_{2j}A^i)
+(A^{i\dag}B_{1j}-\alpha^{-1}B_{1j}A^{i\dag})\right)\right.\right.\nonumber\\
&&\left.\left.\left([H^i,B_{3j}^{\dag}]+(\alpha A^i B_{1j}^{\dag}
-B_{1j}^{\dag}A^i)+(\alpha^{-1}A^{i\dag}B_{2j}^{\dag}-B_{2j}^{\dag}A^{i\dag})
\right)\right.\right.\nonumber\\
&&\left.\left.+\frac{1}{2}\left([B_{1i}, B_{1j}]+(\alpha^{-1}
B_{2i}B_{3j}-\alpha B_{3j}B_{2i})+(\alpha B_{3i}B_{2j}-\alpha^{-1}
B_{2j}B_{3i})\right)\right.\right.\nonumber\\
&&\left.\left.\left([B_{1i}^{\dag},B_{1j}^{\dag}]+(\alpha^{-1} 
B_{2i}^{\dag}B_{3j}^{\dag}-\alpha^{-1}B_{3j}^{\dag}B_{2i}^{\dag})
+(\alpha B_{3i}^{\dag}B_{2j}^{\dag}-\alpha^{-1}B_{2j}^{\dag}B_{3i}^{\dag})
\right)\right.\right.\nonumber\\
&&\left.\left.+\frac{1}{2}\left(\alpha[B_{2i},B_{2j}]
+(B_{3i}B_{1j}-\alpha^{-1}B_{1j}B_{3i})+(\alpha^{-1}B_{1i}B_{3j}
-B_{3j}B_{1i})\right)\right.\right.\nonumber\\
&&\left.\left.\left(\alpha^{-1}[B_{2i}^{\dag},B_{2j}^{\dag}]
+(\alpha B_{3i}^{\dag}B_{1j}^{\dag}-B_{1j}^{\dag}B_{3i}^{\dag})
+(B_{1i}^{\dag}B_{3j}^{\dag}-\alpha B_{3j}^{\dag}B_{1i}^{\dag})
\right)\right.\right.\nonumber\\
&&\left.\left.+\frac{1}{2}\left(\alpha^{-1}[B_{3i},B_{3j}]
+(\alpha B_{1i}B_{2j}-B_{2j}B_{1i})+(B_{2i}B_{1j}-\alpha B_{1i}B_{2j})
\right)\right.\right.\nonumber\\
&&\left.\left.\left(\alpha[B_{3i}^{\dag},B_{3j}^{\dag}]
+(B_{1i}^{\dag}B_{2j}^{\dag}-\alpha^{-1}B_{2j}^{\dag}B_{1i}^{\dag})
+(\alpha^{-1}B_{2i}^{\dag}B_{1j}^{\dag}-B_{1j}^{\dag}B_{2i}^{\dag})
\right)\right.\right.\nonumber\\
&&\left.\left.+\frac{1}{2}\left([B_{1i},B_{1j}^{\dag}]
+[B_{2i},B_{2j}^{\dag}]+[B_{3i},B_{3j}^{\dag}]\right)
\left([B_{1i}^{\dag},B_{1j}]+[B_{2i}^{\dag},B_{2j}]+[B_{3i}^{\dag}
,B_{3j}]\right)\right.\right.\nonumber\\
&&\left.\left.+\frac{1}{2}\left((B_{1i}B_{3j}^{\dag}
-\alpha^{-1}B_{3j}^{\dag}B_{1i}^{\dag})+(B_{2i}B_{1j}^{\dag}
-\alpha^{-1}B_{1j}^{\dag}B_{2i})
+(B_{3i}B_{2j}^{\dag}-\alpha^{-1}B_{2j}^{\dag}B_{3i})\right)
\right.\right.\nonumber\\
&&\left.\left.\left((\alpha B_{1i}^{\dag}B_{3j}-B_{3j}B_{1i}^{\dag})
+(\alpha B_{2i}^{\dag}B_{1j}-B_{1j}B_{2i}^{\dag})+(\alpha B_{3i}^{\dag}
B_{2j}-B_{2j}B_{3i}^{\dag})\right)\right.\right.\nonumber\\
&&\left.\left.+\frac{1}{2}\left((B_{1i}B_{2j}^{\dag}-\alpha 
B_{2j}^{\dag}B_{1i})+(B_{2i}B_{3j}^{\dag}-\alpha B_{3j}^{\dag}
B_{2i})+(B_{3i}B_{1j}^{\dag}-\alpha B_{1j}^{\dag}B_{3i})\right)
\right.\right.\nonumber\\
&&\left.\left.\left((\alpha^{-1}B_{1i}^{\dag}B_{2j}-B_{2j}B_{1i}^{\dag})
+(\alpha^{-1}B_{2i}^{\dag}B_{3j}-B_{3j}B_{2i}^{\dag})+(\alpha^{-1}
B_{3i}^{\dag}B_{1j}-B_{1j}B_{3i}^{\dag})\right)\right\}\right] 
\label{bosonic action}, 
\end{eqnarray} 
\begin{eqnarray}
&S&(\mbox{fermionic action})\nonumber\\
&=&3{\rm Tr}\int d\tau~i\left\{(\bar{\Theta}_{1}{\hat H})
({\dot\Theta}_{1} {\hat H})+(\bar{\Theta}_{1}{\hat A})
({\dot\Theta}_{1}{\hat A}^{\dag}) +(\bar{\Theta}_{1}{\hat A}^{\dag})
({\dot\Theta}_{1}{\hat A})\right.\nonumber\\ 
&&\left.+(\bar{\Theta}_{\omega}{\hat B}_{1\omega}^{\dag})
({\dot \Theta}_{\omega} {\hat B}_{1\omega})+(\bar{\Theta}_{\omega}
{\hat B}_{2\omega}^{\dag})({\dot \Theta}_{\omega} {\hat B}_{2\omega})
+(\bar{\Theta}_{\omega}{\hat B}_{3\omega}^{\dag})({\dot \Theta}_{\omega} 
{\hat B}_{3\omega})\right.\nonumber\\
&&\left.+(\bar{\Theta}_{\omega^{2}}^{\dag}{\hat B}_{1\omega^2}^{\dag})
({\dot \Theta}_{\omega^{2}}{\hat B}_{1\omega^2})
+(\bar{\Theta}_{\omega^{2}}^{\dag}{\hat B}_{2\omega^2}^{\dag})
({\dot \Theta}_{\omega^{2}}{\hat B}_{2\omega^2})
+(\bar{\Theta}_{\omega^{2}}^{\dag}{\hat B}_{3\omega^2}^{\dag})
({\dot \Theta}_{\omega^{2}}{\hat B}_{3\omega^2})\right.\nonumber\\ 
&&\left.+(\bar{\Theta}_{1}{\hat H})([H_{0},(\Theta_{1}{\hat H})] 
+[A_{0},(\Theta_{1}{\hat A}^{\dag})]+[A_{0}^{\dag},(\Theta_{1}{\hat A})]) 
\right.\nonumber\\
&&\left.+(\bar{\Theta}_{1}{\hat A})([H_{0},(\Theta_{1}{\hat A}^{\dag})]
+[A_{0},(\Theta_{1}{\hat A})]+[A_{0}^{\dag},(\Theta_{1}{\hat H})]) 
\right.\nonumber\\
&&\left.+(\bar{\Theta}_{1}{\hat A}^{\dag})([H_{0},(\Theta_{1} 
{\hat A})]+[A_{0},(\Theta_{1}{\hat H})]+[A_{0}^{\dag},(\Theta_{1} 
{\hat A}^{\dag})])\right.\nonumber\\
&&\left.+(\bar{\Theta}_{\omega}{\hat B}_{1\omega}^{\dag})
\left([H_{0},(\Theta_{\omega} {\hat B}_{1\omega})]+(A_0(\Theta_{\omega}
{\hat B}_{3\omega})-\alpha(\Theta_{\omega}{\hat B}_{3\omega})A_0)
+(A_0^{\dag}(\Theta_{\omega}{\hat B}_{2\omega})-\alpha^{-1}
(\Theta_{\omega}{\hat B}_{2\omega})A_0^{\dag})\right)\right.\nonumber\\
&&\left.+(\bar{\Theta}_{\omega}{\hat B}_{2\omega}^{\dag})
\left([H_{0},(\Theta_{\omega} {\hat B}_{2\omega})]
+(A_0(\Theta_{\omega}{\hat B}_{1\omega})-\alpha(\Theta_{\omega}
{\hat B}_{1\omega})A_0)+(A_0^{\dag}(\Theta_{\omega}
{\hat B}_{3\omega})-\alpha^{-1}(\Theta_{\omega}{\hat B}_{3\omega})
A_0^{\dag})\right)\right.\nonumber\\
&&\left.+(\bar{\Theta}_{\omega}{\hat B}_{3\omega}^{\dag})
\left([H_{0},(\Theta_{\omega} {\hat B}_{3\omega})]
+(A_0(\Theta_{\omega}{\hat B}_{2\omega})-\alpha(\Theta_{\omega}
{\hat B}_{2\omega})A_0)+(A_0^{\dag}(\Theta_{\omega}{\hat B}_{1\omega})
-\alpha^{-1}(\Theta_{\omega}{\hat B}_{1\omega})A_0^{\dag})
\right)\right.\nonumber\\
&&\left.+(\bar{\Theta}_{\omega^2}{\hat B}_{1\omega^2}^{\dag})\left([H_{0},
(\Theta_{\omega^2} {\hat B}_{1\omega^2})]+(A_0(\Theta_{\omega^2}
{\hat B}_{3\omega^2})-\alpha^{-1}(\Theta_{\omega^2}{\hat B}_{3\omega^2})
A_0)\right.\right.\nonumber\\
&&\left.\left.+(A_0^{\dag}(\Theta_{\omega}{\hat B}_{2\omega^2})-\alpha(
\Theta_{\omega^2}{\hat B}_{2\omega^2})A_0^{\dag})\right)\right.\nonumber\\
&&\left.+(\bar{\Theta}_{\omega^2}{\hat B}_{2\omega^2}^{\dag})
\left([H_{0},(\Theta_{\omega^2} {\hat B}_{2\omega^2})]
+(A_0(\Theta_{\omega^2}{\hat B}_{1\omega^2})-\alpha^{-1}
(\Theta_{\omega^2}{\hat B}_{1\omega^2})A_0)\right.\right.\nonumber\\
&&\left.\left.+(A_0^{\dag}(\Theta_{\omega^2}{\hat B}_{3\omega^2})
-\alpha(\Theta_{\omega^2}{\hat B}_{3\omega^2})A_0^{\dag})
\right)\right.\nonumber\\
&&\left.+(\bar{\Theta}_{\omega^2}{\hat B}_{3\omega^2}^{\dag})
\left([H_{0},(\Theta_{\omega^2} {\hat B}_{3\omega^2})]
+(A_0(\Theta_{\omega^2}{\hat B}_{2\omega^2})-\alpha^{-1}
(\Theta_{\omega^2}{\hat B}_{2\omega^2})A_0)\right.\right.\nonumber\\
&&\left.\left.+(A_0^{\dag}(\Theta_{\omega^2}{\hat B}_{1\omega^2})
-\alpha(\Theta_{\omega^2}{\hat B}_{1\omega^2})A_0^{\dag})
\right)\right\}\nonumber\\
&+&3{\rm Tr}\int d\tau~R\left\{(\bar{\Theta}_{1}{\hat H})\Gamma_{i} 
([H^{i},(\Theta_{1}{\hat H})]+[A^{i},(\Theta_{1}{\hat A}^{\dag})] 
+[A^{i\dag},(\Theta_{1}{\hat A})])\right.\nonumber\\ 
&&\left.+(\bar{\Theta}_{1}{\hat A})\Gamma_{i} ([H^{i},(\Theta_{1}
{\hat A}^{\dag})]+[A^{i},(\Theta_{1}{\hat A})] +[A^{i\dag},
(\Theta_{1}{\hat H})])\right.\nonumber\\ 
&&\left.+(\bar{\Theta}_{1}{\hat A}^{\dag})\Gamma_{i} ([H^{i},
(\Theta_{1}{\hat A})]+[A^{i},(\Theta_{1}{\hat H})] +[A^{i\dag},
(\Theta_{1}{\hat A}^{\dag})])\right.\nonumber\\ 
&&\left.+(\bar{\Theta}_{\omega}{\hat B}_{1\omega}^{\dag})\Gamma_{i}
\left( [H^{i},(\Theta_{\omega}{\hat B}_{1\omega})]
+(A^i(\Theta_{\omega}{\hat B}_{3\omega})-\alpha(\Theta_{\omega}
{\hat B}_{3\omega})A^i)+(A^{i\dag}(\Theta_{\omega}{\hat B}_{2\omega})
-\alpha^{-1}(\Theta_{\omega}{\hat B}_{2\omega})A^{i\dag})
\right)\right.\nonumber\\
&&\left.+(\bar{\Theta}_{\omega}{\hat B}_{2\omega}^{\dag})\Gamma_{i}
\left( [H^{i},(\Theta_{\omega}{\hat B}_{2\omega})]
+(A^i(\Theta_{\omega}{\hat B}_{1\omega})-\alpha(\Theta_{\omega}
{\hat B}_{1\omega})A^i)+(A^{i\dag}(\Theta_{\omega}{\hat B}_{3\omega})
-\alpha^{-1}(\Theta_{\omega}{\hat B}_{3\omega})A^{i\dag})
\right)\right.\nonumber\\
&&\left.+(\bar{\Theta}_{\omega}{\hat B}_{3\omega}^{\dag})\Gamma_{i}
\left( [H^{i},(\Theta_{\omega}{\hat B}_{3\omega})]+(A^i(\Theta_{\omega}
{\hat B}_{2\omega})-\alpha(\Theta_{\omega}{\hat B}_{2\omega})A^i)
+(A^{i\dag}(\Theta_{\omega}{\hat B}_{1\omega})-\alpha^{-1}
(\Theta_{\omega}{\hat B}_{1\omega})A^{i\dag})\right)\right.\nonumber\\
&&\left.+(\bar{\Theta}_{\omega^2}{\hat B}_{1\omega^2}^{\dag})\Gamma_{i}
\left( [H^{i},(\Theta_{\omega^2}{\hat B}_{1\omega^2})]
+(A^i(\Theta_{\omega^2}{\hat B}_{3\omega^2})-\alpha^{-1}
(\Theta_{\omega^2}{\hat B}_{3\omega^2})A^i)\right.\right.\nonumber\\
&&\left.\left.+(A^{i\dag}(\Theta_{\omega^2}
{\hat B}_{2\omega^2})-\alpha(\Theta_{\omega^2}{\hat B}_{2\omega^2})
A^{i\dag})\right)\right.\nonumber\\
&&\left.+(\bar{\Theta}_{\omega^2}{\hat B}_{2\omega^2}^{\dag})\Gamma_{i}
\left( [H^{i},(\Theta_{\omega^2}{\hat B}_{2\omega^2})]
+(A^i(\Theta_{\omega^2}{\hat B}_{1\omega^2})-\alpha^{-1}
(\Theta_{\omega^2}{\hat B}_{1\omega^2})A^i)\right.\right.\nonumber\\
&&\left.\left.+(A^{i\dag}
(\Theta_{\omega^2}{\hat B}_{3\omega^2})-\alpha(\Theta_{\omega^2}
{\hat B}_{3\omega^2})A^{i\dag})\right)\right.\nonumber\\
&&\left.+(\bar{\Theta}_{\omega^2}{\hat B}_{3\omega^2}^{\dag})\Gamma_{i}
\left( [H^{i},(\Theta_{\omega^2}{\hat B}_{3\omega^2})]
+(A^i(\Theta_{\omega^2}{\hat B}_{2\omega^2})-\alpha^{-1}
(\Theta_{\omega^2}{\hat B}_{2\omega^2})A^i)\right.\right.\nonumber\\
&&\left.\left.+(A^{i\dag}(\Theta_{\omega^2}
{\hat B}_{1\omega^2})-\alpha(\Theta_{\omega^2}{\hat B}_{1\omega^2})
A^{i\dag})\right)\right.\nonumber\\
&&\left.+(\bar{\Theta}_{1}{\hat H})b_{i+1}\left([B_{1i},
(\Theta_{\omega^{2}}{\hat B}_{1\omega^2})]+\alpha^{-1}[B_{2i},
(\Theta_{\omega^2}B_{3\omega^2})]+\alpha[B_{3i},(\Theta_{\omega^2}
{\hat B}_{2\omega^2})]\right)\right.\nonumber\\
&&\left.+(\bar{\Theta}_{1}{\hat A})b_{i+1}\left((\alpha B_{1i}
(\Theta_{\omega^2}{\hat B}_{2\omega^2})-(\Theta_{\omega^2}
{\hat B}_{2\omega^2})B_{1i})+(B_{2i}(\Theta_{\omega^2}
{\hat B}_{1\omega^2})-\alpha^{-1}(\Theta_{\omega^2}{\hat B}_{1\omega^2})
B_{2i})\right.\right.\nonumber\\
&&\left.\left.+(\alpha^{-1}B_{3i}(\Theta_{\omega^2}{\hat B}_{3\omega^2})
-\alpha(\Theta_{\omega^2}{\hat B}_{3\omega^2})B_{3i})\right)\right.\nonumber\\
&&\left.+(\bar{\Theta}_{1}{\hat A}^{\dag})b_{i+1}\left((\alpha^{-1} 
B_{1i}(\Theta_{\omega^2}{\hat B}_{3\omega^2})-(\Theta_{\omega^2}
{\hat B}_{3\omega^2})B_{1i})+(\alpha B_{2i}(\Theta_{\omega^2}
{\hat B}_{2\omega^2})-\alpha^{-1}(\Theta_{\omega^2}{\hat B}_{2\omega^2})
B_{2i})\right.\right.\nonumber\\
&&\left.\left.+(B_{3i}(\Theta_{\omega^2}{\hat B}_{1\omega^2})
-\alpha(\Theta_{\omega^2}{\hat B}_{1\omega^2})B_{3i})\right)\right.\nonumber\\
&&\left.+(\bar{\Theta}_{\omega}B_{1\omega}^{\dag})b_{i+1}
\left([B_{1i},(\Theta_1{\hat H})]+(\alpha^{-1}B_{2i}
(\Theta_1{\hat A}^{\dag})-(\Theta_1{\hat A}^{\dag})B_{2i})
+(\alpha B_{3i}(\Theta_{1}{\hat A})-(\Theta_1{\hat A})B_{3i})
\right)\right.\nonumber\\
&&\left.+(\bar{\Theta}_{\omega}B_{2\omega}^{\dag})b_{i+1}\left(
(\alpha B_{1i}(\Theta_1{\hat A})-(\Theta_1{\hat A})B_{1i})
+[B_{2i},(\Theta_1{\hat H}]+(\alpha^{-1} B_{3i}(\Theta_{1}{\hat A}^{\dag})
-(\Theta_1{\hat A}^{\dag})B_{3i})\right)\right.\nonumber\\
&&\left.+(\bar{\Theta}_{\omega}B_{3\omega}^{\dag})b_{i+1}\left(
(\alpha^{-1}B_{1i}(\Theta_1{\hat A})-(\Theta_1{\hat A})B_{1i})
+(\alpha B_{2i}(\Theta_1{\hat A})-(\Theta_1{\hat A}^{\dag})B_{2i})
+[B_{3i},(\Theta_{1}{\hat H})]\right)\right.\nonumber\\
&&\left.+(\bar{\Theta}_{\omega^2}B_{1\omega^2}^{\dag})b_{i+1}
\left([B_{1i},(\Theta_{\omega}{\hat B}_{1\omega})]+(\alpha^{-1}B_{2i}
(\Theta_\omega{\hat B}_{3\omega})-\alpha(\Theta_\omega{\hat B}_{3\omega})
B_{2i})\right.\right.\nonumber\\
&&\left.\left.+(\alpha B_{3i}(\Theta_{\omega}{\hat B}_{2\omega})
-\alpha^{-1}(\Theta_\omega{\hat B}_{2\omega})B_{3i})\right)\right.\nonumber\\
&&\left.+(\bar{\Theta}_{\omega^2}B_{2\omega^2}^{\dag})b_{i+1}
\left((\alpha B_{1i}(\Theta_{\omega}{\hat B}_{2\omega})-(\Theta_{\omega}
{\hat B}_{2\omega})B_{1i})+(B_{2i}(\Theta_{\omega}{\hat B}_{1\omega}
-\alpha(\Theta_{\omega}{\hat B}_{1\omega})B_{2i})\right.\right.\nonumber\\
&&\left.\left.+\alpha^{-1}[B_{3i},(\Theta_{\omega}{\hat B}_{3\omega})]
\right)\right.\nonumber\\
&&\left.+(\bar{\Theta}_{\omega^2}B_{3\omega^2}^{\dag})b_{i+1}
\left((\alpha^{-1}B_{1i}(\Theta_{\omega}{\hat B}_{3\omega})
-(\Theta_{\omega}{\hat B}_{3\omega})B_{1i})+\alpha [B_{2i},
(\Theta_\omega{\hat B}_{2\omega})]\right.\right.\nonumber\\
&&\left.\left.+(B_{3i}(\Theta_{\omega}{\hat B}_{1\omega})
-\alpha^{-1}(\Theta_{\omega}{\hat B}_{1\omega})B_{3i})\right)\right.\nonumber\\
&&\left.+(\bar{\Theta}_{1}{\hat H})b_{i+1}^{\dag}\left([B_{1i}^{\dag},
(\Theta_{\omega}{\hat B}_{1\omega})]+[B_{2i}^{\dag},
(\Theta_{\omega}B_{2\omega})]+[B_{3i}^{\dag},
(\Theta_{\omega}{\hat B}_{3\omega})]\right)\right.\nonumber\\
&&\left.+(\bar{\Theta}_{1}{\hat A}^{\dag})b_{i+1}^{\dag}\left(
(\alpha^{-1} B_{1i}^{\dag}(\Theta_{\omega}{\hat B}_{2\omega})
-(\Theta_{\omega}{\hat B}_{2\omega})B_{1i}^{\dag})+(\alpha^{-1}
B_{2i}^{\dag}(\Theta_{\omega}{\hat B}_{3\omega})-(\Theta_{\omega}
{\hat B}_{3\omega})B_{2i}^{\dag})\right.\right.\nonumber\\
&&\left.\left.+(\alpha^{-1} B_{3i}^{\dag}
(\Theta_{\omega}{\hat B}_{1\omega})-(\Theta_{\omega}{\hat B}_{1\omega})
B_{3i}^{\dag})\right)\right.\nonumber\\
&&\left.+(\bar{\Theta}_{1}{\hat A})b_{i+1}^{\dag}\left((\alpha 
B_{1i}^{\dag}(\Theta_{\omega}{\hat B}_{3\omega})-(\Theta_{\omega}
{\hat B}_{3\omega})B_{1i}^{\dag})+(\alpha B_{2i}^{\dag}(\Theta_{\omega}
{\hat B}_{1\omega})-(\Theta_{\omega}{\hat B}_{1\omega})B_{2i}^{\dag})
\right.\right.\nonumber\\
&&\left.\left.+(\alpha B_{3i}^{\dag}(\Theta_{\omega}{\hat B}_{2\omega})
-(\Theta_{\omega}{\hat B}_{2\omega})B_{3i}^{\dag})\right)\right.\nonumber\\
&&\left.+(\bar{\Theta}_{\omega}B_{1\omega}^{\dag})b_{i+1}^{\dag}
\left([B_{1i}^{\dag},(\Theta_{\omega^2}{\hat B}_{1\omega^2})]
+(B_{2i}^{\dag}(\Theta_{\omega^2}{\hat B}_{2\omega^2})-\alpha^{-1}
(\Theta_{\omega^2}{\hat B}_{2\omega^2})B_{2i}^{\dag})\right.\right.\nonumber\\
&&\left.\left.+(B_{3i}^{\dag}(\Theta_{\omega^2}{\hat B}_{3\omega^2})
-\alpha(\Theta_{\omega^2}{\hat B}_{3\omega^2})B_{3i}^{\dag})\right)
\right.\nonumber\\
&&\left.+(\bar{\Theta}_{\omega}B_{2\omega}^{\dag})b_{i+1}^{\dag}
\left((\alpha^{-1} B_{1i}^{\dag}(\Theta_{\omega^2}{\hat B}_{2\omega^2})
-(\Theta_{\omega^2}{\hat B}_{2\omega^2})B_{1i}^{\dag})+\alpha^{-1}
[B_{2i}^{\dag},(\Theta_{\omega^2}{\hat B}_{3\omega^2})]
\right.\right.\nonumber\\
&&\left.\left.+(\alpha^{-1} B_{3i}^{\dag}(\Theta_{\omega^2}
{\hat B}_{1\omega^2})-\alpha (\Theta_{\omega^2}{\hat B}_{1\omega^2})
B_{3i}^{\dag})\right)\right.\nonumber\\
&&\left.+(\bar{\Theta}_{\omega}B_{3\omega}^{\dag})b_{i+1}^{\dag}
\left((\alpha B_{1i}^{\dag}(\Theta_{\omega^2}{\hat B}_{3\omega^2})
-(\Theta_{\omega^2}{\hat B}_{3\omega^2})B_{1i}^{\dag})
+(\alpha B_{2i}^{\dag}(\Theta_{\omega^2}{\hat B}_{1\omega^2})
-\alpha^{-1}(\Theta_{\omega^2}{\hat B}_{1\omega^2})B_{2i}^{\dag})
\right.\right.\nonumber\\
&&\left.\left.+\alpha[B_{3i}^{\dag},(\Theta_{\omega^2}
{\hat B}_{2\omega^2})]\right)\right.\nonumber\\
&&\left.+(\bar{\Theta}_{\omega^2}B_{1\omega^2}^{\dag})b_{i+1}^{\dag}
\left([B_{1i}^{\dag},(\Theta_{1}{\hat H})]+(B_{2i}^{\dag}
(\Theta_1{\hat A})-\alpha(\Theta_1{\hat A})B_{2i}^{\dag})+(B_{3i}^{\dag}
(\Theta_1{\hat A}^{\dag})-\alpha^{-1}(\Theta_1{\hat A}^{\dag})
B_{3i}^{\dag})\right)\right.\nonumber\\
&&\left.+(\bar{\Theta}_{\omega^2}B_{2\omega^2}^{\dag})b_{i+1}^{\dag}
\left((\alpha^{-1} B_{1i}^{\dag}(\Theta_1{\hat A})-(\Theta_1{\hat A})
B_{1i}^{\dag})+(\alpha^{-1}B_{2i}^{\dag}((\Theta_1{\hat A}^{\dag})
-\alpha (\Theta_1{\hat A}^{\dag})B_{2i}^{\dag})+\alpha^{-1}
[B_{3i}^{\dag},(\Theta{\hat H})]\right)\right.\nonumber\\
&&\left.+(\bar{\Theta}_{\omega^2}B_{3\omega^2}^{\dag})b_{i+1}^{\dag}
\left((\alpha B_{1i}^{\dag}(\Theta_1{\hat A}^{\dag})-(\Theta_1{\hat A}^{\dag})
B_{1i}^{\dag})+\alpha [B_{2i}^{\dag},(\Theta_1{\hat H})]+(\alpha 
B_{3i}^{\dag}(\Theta_1{\hat A})-\alpha^{-1}(\Theta_1{\hat A})B_{3i}^{\dag})
\right)\right\} \label{fermionic action},
\end{eqnarray}
\begin{eqnarray}
&S&(\mbox{ghost action})\nonumber\\
&=&-\int d\tau {\rm Tr}\left(\bar{C}\partial^0 D_0 C-\bar{C}
[X_{cl}^i ,[X^i ,C]]\right)\nonumber\\
&=&-3{\rm Tr}\int d\tau
\left[C_{gh}^{\dag}(\partial_0)^2 C_{gh}+A_{gh}^{\dag}
(\partial_0)^2 A_{gh}+A_{gh}(\partial_0)^2 A_{gh}^{\dag}\right.\nonumber\\
&&\left.-i\left(C_{gh}^{\dag}[H_0,C_{gh}]+C_{gh}^{\dag}[A_{0\dag},A_{gh}]
+C_{gh}^{\dag}[A_0,A_{gh}^{\dag}]\right.\right.\nonumber\\
&&\left.\left.+A_{gh}^{\dag}[H_0,A_{gh}]+A_{gh}^{\dag}[A_{0}^{\dag},
A_{gh}^{\dag}]+A_{gh}^{\dag}[A_0,C_{gh}]\right.\right.\nonumber\\
&&\left.\left.+A_{gh}[H_0,A_{gh}^{\dag}]+A_{gh}[A_{0}^{\dag},C_{gh}]
+A_{gh}[A_0,A_{gh}^{\dag}]\right)\right.\nonumber\\
&&\left.-\left(C_{gh}^{\dag}[H_{cl}^i ,[H^i ,C_{gh}]]
+ C_{gh}^{\dag}[H_{cl}^i ,[A^{i\dag},A_{gh}]]+C_{gh}^{\dag} [H_{cl}^i ,
[A^i ,A_{gh}^{\dag}]]\right.\right.\nonumber\\ 
&&\left.\left.+A_{gh}^{\dag}[H_{cl}^i ,[H^i ,A_{gh}]]+A_{gh}^{\dag}
[H_{cl}^i,[A^{i\dag},A_{gh}]]+A_{gh}^{\dag}[H_{cl}^i ,[A^i ,C_{gh}]]
\right.\right.\nonumber\\ 
&&\left.\left.+A_{gh}[H_{cl}^i ,[H^i ,A_{gh}^{\dag}]]+A_{gh}[H_{cl}^{i\dag}, 
[A^{i\dag},C_{gh}]]+A_{gh}[H_{cl}^i ,[A^i ,A_{gh}]]\right)\right.\nonumber\\
&&\left.+\frac{1}{2}\left\{
C_{gh}^{\dag}\left([B_{1icl},[B_{1i}^{\dag},C_{gh}]]\right.\right.\right.
\nonumber\\
&&\quad\left.\left.\left.
+B_{1icl}(B_{2i}^\dag A_{gh}-\alpha A_{gh}B_{2i}^{\dag})
-(B_{2i}^{\dag}A_{gh}-\alpha A_{gh}B_{2i}^{\dag})B_{1icl}
\right.\right.\right.\nonumber\\
&&\quad\left.\left.\left. +B_{1icl}(B_{3i}^\dag A_{gh}-\alpha^{-1} 
A_{gh} B_{3i}^{\dag})
-(B_{3i}^\dag A_{gh}-\alpha^{-1}A_{gh}B_{3i}^{\dag})B_{1icl}\right.\right.
\right.\nonumber\\
&&\quad\left.\left.\left.+[B_{1icl}^{\dag},[B_{1i},C_{gh}]]+B_{1icl}^{\dag}
(\alpha^{-1}B_{2i}
A_{gh}^{\dag}-A_{gh}^{\dag}B_{2i})-(\alpha^{-1}B_{2i}A_{gh}^{\dag}
-A_{gh}^{\dag}B_{2i})B_{1icl}^{\dag}\right.\right.\right.\nonumber\\
&&\quad\left.\left.\left.+B_{1icl}^{\dag}(\alpha B_{3i}A_{gh}-A_{gh}B_{3i})
-(\alpha B_{3i}A_{gh}-A_{gh}B_{3i})B_{1icl}^{\dag}\right)\right.\right.
\nonumber\\
&&\left.\left.+A_{gh}^{\dag}\left(B_{1icl}(B_{1i}^\dag A_{gh}-\alpha A_{gh}
B_{1i}^{\dag})-(\alpha^{-1}B_{1i}^{\dag}A_{gh}-A_{gh}B_{1i}^{\dag})B_{1icl}
\right.\right.\right.\nonumber\\
&&\quad\left.\left.\left.
+B_{1icl}(B_{2i}^{\dag}A_{gh}-\alpha^{-1}A_{gh}B_{2i}^{\dag})
-(\alpha^{-1}B_{2i}^{\dag}A_{gh}-\alpha A_{gh}B_{2i}^{\dag})B_{1icl}
\right.\right.\right.\nonumber\\
&&\quad\left.\left.\left.+B_{1icl}[B_{3i}^{\dag},C_{gh}]-\alpha^{-1}[B_{3i}^{\dag},
C_{gh}]B_{1icl}\right.\right.\right.\nonumber\\
&&\quad\left.\left.\left.+B_{1icl}^{\dag}(B_{1i}A_{gh}-\alpha^{-1}A_{gh}B_{1i})
-(\alpha B_{1i}A_{gh}-A_{gh}B_{1i})B_{1icl}^{\dag}
\right.\right.\right.\nonumber\\
&&\quad\left.\left.\left.
+\alpha^{-1}B_{1icl}^{\dag}
[B_{2i},C_{gh}]-[B_{2i},C_{gh}]B_{1icl}^{\dag}\right.\right.\right.\nonumber\\
&&\quad\left.\left.\left.+B_{1icl}^{\dag}(\alpha B_{3i}A_{gh}-\alpha^{-1}
A_{gh}^{\dag}
B_{3i})-(\alpha^{-1}B_{3i}A_{gh}^{\dag}-A_{gh}^{\dag}B_{3i})
B_{1icl}^{\dag}\right)\right.\right.\nonumber\\
&&\left.\left.+A_{gh}\left(B_{1icl}(B_{1i}^{\dag}A_{gh}^{\dag}-\alpha^{-1}
A_{gh}^{\dag}B_{1i}^{\dag})-(\alpha B_{1i}^{\dag}A_{gh}^{\dag}-A_{gh}
B_{1i}^{\dag})B_{1icl}
\right.\right.\right.\nonumber\\
&&\quad\left.\left.\left.
+B_{1icl}[B_{2i}^{\dag},C_{gh}]-\alpha[B_{2i}^{\dag},C_{gh}]B_{1icl}
\right.\right.\right.\nonumber\\
&&\quad\left.\left.\left.+B_{1icl}(B_{3i}^{\dag}A_{gh}-\alpha A_{gh}B_{3i}^{\dag})
-(\alpha B_{3i}^{\dag}A_{gh}-\alpha^{-1}A_{gh}B_{3i}^{\dag})B_{1icl}
\right.\right.\right.\nonumber\\
&&\quad\left.\left.\left.+B_{1icl}^{\dag}(B_{1i}A_{gh}^{\dag}-\alpha A_{gh}^{\dag}
B_{1i})-(\alpha^{-1}B_{1i}A_{gh}^{\dag}-A_{gh}^{\dag}B_{1i})B_{1icl}^{\dag}
\right.\right.\right.\nonumber\\
&&\quad\left.\left.\left.
+B_{1icl}^{\dag}(\alpha^{-1}B_{2i}A_{gh}-\alpha A_{gh}B_{2i})
-(\alpha B_{2i}A_{gh}-A_{gh}B_{2i})B_{1icl}^{\dag}\right.\right.\right.
\nonumber\\
&&\quad\left.\left.\left.+\alpha B_{1icl}^{\dag}[B_{3i},C_{gh}]-[B_{3i},C_{gh}]
B_{1icl}^{\dag}\right)\right\}\right] \label{ghost action}.
\end{eqnarray}}




\end{document}